\documentclass[12pt]{article}
\usepackage[dvips]{color}
\usepackage{epsfig}
\usepackage{amsmath}
\usepackage{graphicx}
\def\Box{\hbox{$\rlap{$\sqcup$}\sqcap$}}
\begin{document}
\title{\bf Non-minimal Maxwell-Modified Gauss-Bonnet Cosmologies: Inflation and Dark Energy  }
\author{J. Sadeghi $^{a}$\thanks{Email: pouriya@ipm.ir}\hspace{1mm}
, M. R. Setare $^{b}$ \thanks{Email: rezakord@ipm.ir}\hspace{1mm}
and
 A. Banijamali $^{a}$\thanks{Email: abanijamali@umz.ac.ir}\hspace{1mm} \\
$^a$ {\small {\em  Sciences Faculty, Department of Physics, Mazandaran University,}}\\
{\small {\em P .O .Box 47415-416, Babolsar, Iran}}\\
$^{b}${\small {\em Department of Science, Payame Noor University, Bijar, Iran }}\\
} \maketitle
\begin{abstract}
In this paper we show that power-law inflation can be realized in
non-minimal gravitational coupling of electromagnetic field with a
general function of Gauss-Bonnet invariant. Such a non-minimal
coupling may appear due to quantum corrections. We also consider
modified Maxwell-$F(G)$ gravity in which non-minimal coupling
between electromagnetic field and $f(G)$ occur in the framework of
modified Gauss-Bonnet gravity. It is shown that inflationary
cosmology and late-time accelerated expansion of the universe are
possible in such a theory.

 \noindent
\hspace{0.35cm}


\end{abstract}
\section{Introduction}
Cosmological observations indicate that there are two periods of
accelerated expansion in our universe: cosmic inflation in the early
universe and acceleration in the current expansion of the universe
[1-4]. \\
In order to explain the late-time acceleration of the universe, one
needs to introduce a negative pressure component which is called
dark energy (DE). In this direction we can consider field models of
dark energy. The field models that have been discussed widely in the
literature consider a cosmological constant \cite{cosmo}, a
canonical scalar field (quintessence) \cite{quint}, a phantom field,
that is a scalar field with a negative sign of the kinetic term
\cite{phant,phantBigRip}, or the combination of quintessence and
phantom in a unified model named quintom \cite{quintom}.\\
An alternative approach to explain DE is the modification of general
relativity (GR) \cite{{c7},{c8},{c9},{c10},{c11},{c12}}, and in the
simplest way adding an arbitrary function of Ricci scalar to the
Einstein-Hilbert action, what is well known
as $f(R)$ gravity \cite{{c13},{c14}}.\\
Another modification of GR is modified Gauss-Bonnet gravity [15],
which is obtain by inserting a general function of Gauss-Bonnet
invariant, $f(G)$, in the GR action.\\
In the other side non-minimal coupling between the Ricci scalar and
matter Lagrangian can be seen as the source of inflation and current
acceleration expansion of the universe \cite{{c16},{c17}}. Such a
non-minimal coupling between $f(R)$/$f(G)$ gravity and kinetic part
of Lagrangian of a massless scalar field has been investigated in
Ref. \cite{c18}. Non-minimal coupling of a viable $f(R)$ gravity
with electromagnetism Lagrangian and cosmological consequences of
this model about inflation and late-time acceleration has been
considered in \cite{c19}. Also cosmology in non-minimal non-Abelian
gauge theory (Yang-Mills theory), in which the non-Abelian gauge
field couples to $f(R)$ gravity has been explored in Ref.
\cite{c20}. Furthermore, it is shown that both inflation and late
time accelerated expansion of the universe can be realized in
non-minimal vector model in the framework of modified gravity
\cite{c20}. In addition, the conditions for the non-minimal
gravitational coupling of electromagnetic field in order that
finite-time singularities can not appear have been investigated in
Ref. \cite{c21}. Ref. \cite{c22}, has been considered $F(R)$ gravity
coupled to non-linear electrodynamics. The criteria for the validity
of non-minimal coupling between scalar curvature and matter
Lagrangian have been studied in
Refs. \cite{{c23},{c24},{c25}}.\\
In the present work following the Ref. \cite{c19}, we consider
inflation and late time acceleration of the universe in non-minimal
electromagnetism in which the electromagnetism Lagrangian couples to
the arbitrary function of Gauss-Bonnet invariant, $f(G)$.
Additionally we use the procedure of Ref. \cite{c26} in analyzing
the electromagnetism in large scales. We show that power-law
inflation can be realized in non-minimally coupled electromagnetism
Lagrangian with modified Gauss-Bonnet gravity in the framework of
GR. Furthermore, we study inflation and late time accelerated
expansion of the universe in non-minimal Maxwell-$f(G)$ gravity in
the framework of modified Gauss-Bonnet gravity. We use the proposal
of Ref. \cite{c27}, for a viable $f(G)$ gravity and note that in
$f(G)$ gravity
, there are no problems with the Newton law and instabilities \cite{c28}.\\
An outline of this paper is as follows. In section 2 we examine
power-law inflation in a non-minimally coupled Maxwell field with
$f(G)$ gravity in GR framework. In section 3 we show that both
inflation and late-time cosmic acceleration can be realized in a
model of non-minimal gravitational coupling of the Maxwell field in
a modified Gauss-Bonnet gravity proposed in Ref. \cite{c27}. Section
4 is devoted to conclusion.

\section{Power-law inflation in general relativity}

We start with the following action:

\begin{eqnarray}
S=\int d^{4}x
\sqrt{-g}\Big[\frac{1}{2\kappa^{2}}R-\frac{1}{4}F_{\mu\nu}F^{\mu\nu}-\frac{1}{4}f(G)F_{\mu\nu}F^{\mu\nu}\Big],
\end{eqnarray}
where the second term in (1) is the Lagrangian of Maxwell theory and
the third term represent non-minimal coupling between
electromagnetism Lagrangian and modified Gauss-Bonnet gravity.
$F_{\mu\nu}=\partial_{\mu}A_{\nu}-\partial_{\nu}A_{\mu}$, is the
electromagnetic field strength and $A_{\mu}$ is the $U(1)$ gauge
field. Without the third term, (1) correspond to the usual
Einstein-Maxwell theory. Varying the action (1) with respect to
$A_{\mu}$ lead to following equations of motion for $U(1)$ gauge
field
\begin{equation}
\frac{1}{\sqrt{-g}}\partial_{\mu}\Big[\sqrt{-g}\big(1+f(G)\big)F^{\mu\nu}\Big]=0.
\end{equation}
In addition to the field equations (2), by varying (1) with respect
to $g_{\mu\nu}$,  we obtain
\begin{eqnarray}
0=\frac{1}{\sqrt{-g}}\frac{\delta S}{\delta
g^{\mu\nu}}=\frac{1}{2\kappa^{2}}(\frac{1}{2}g_{\mu\nu}R-R_{\mu\nu})+T_{\mu\nu}^{eff}.
\end{eqnarray}
Here the effective energy momentum tensor $T_{\mu\nu}^{eff}$ is
defined by
\begin{eqnarray}
T_{\mu\nu}^{eff}&=&\frac{1}{2}(1+f(G))\Big(g^{\gamma\delta}F_{\mu\delta}F_{\nu\gamma}-\frac{1}{4}
g_{\mu\nu}F_{\gamma\delta}F^{\gamma\delta}\Big)
+\frac{1}{2}\Big\{f'(G)F_{\gamma\delta}F^{\gamma\delta}R\,R_{\mu\nu}
\nonumber\\
&-&2f'(G)F_{\gamma\delta}F^{\gamma\delta}R_{\mu}\,^{\rho}R_{\nu\rho}+
f'(G)F_{\gamma\delta}F^{\gamma\delta}R_{\mu\rho\sigma\lambda}R_{\nu}\,^{\rho\sigma\lambda}
+2f'(G)F_{\gamma\delta}F^{\gamma\delta}R_{\mu\rho\sigma\nu}R^{\rho\sigma}
\nonumber\\ &+& [(g_{\mu\nu}\Box-\nabla_{\mu}\nabla_{\nu})
f'(G)F_{\gamma\delta}F^{\gamma\delta}]R +2[\nabla^{\rho}\nabla_{\mu}
(f'(G)F_{\gamma\delta}F^{\gamma\delta})]R_{\nu\rho}\nonumber\\
&+&2[\nabla^{\rho}\nabla_{\nu}
(f'(G)F_{\gamma\delta}F^{\gamma\delta})]R_{\mu\rho} +2[\Box
(f'(G)F_{\gamma\delta}F^{\gamma\delta})]R_{\mu\nu}\nonumber\\
&-&2[g_{\mu\nu}\nabla^{\lambda}
\nabla^{\rho}(f'(G)F_{\gamma\delta}F^{\gamma\delta})]
R_{\lambda\rho}+2[\nabla^{\rho}
\nabla^{\lambda}(f'(G)F_{\gamma\delta}F^{\gamma\delta})]R_{\mu\rho\nu\lambda}\Big\},
\end{eqnarray}
where $f'(G)=\frac{df(G)}{dG}$ and
$\Box=g^{\mu\nu}\nabla_{\mu}\nabla_{\nu}$, is the d'Alembertian
operator.\\
In a flat Friedmann-Robertson-Walker (FRW) space-time with the
metric
\begin{eqnarray}
ds^{2}=-dt^{2}+a^{2}(t)(dr^{2}+r^{2}d\Omega^{2}),
\end{eqnarray}
the  components of  Ricci tensor $R_{\mu\nu}$ and Ricci scalar $R$,
are given by
\begin{equation}
R_{00}=-3\big(\dot{H}+H^{2}\big),\,\,\,\,\,R_{ij}=a^{2}(t)\big(\dot{H}+3H^{2}\big)\delta_{ij},\,\,\,\,\,
R=6\big(\dot{H}+2H^{2}\big),
\end{equation}
where $H=\frac{\dot{a}(t)}{a(t)}$ is the Hubble parameter and $a(t)$
is the scale factor. Also Gauss- Bonnet invariant in this background
is
\begin{equation}
 G=24\big(\dot{H}H^{2}+H^{4}\big).
\end{equation}
The $(0,0)$ component and sum of $(i,i)$ components of equation (3)
in FRW space- time have the following forms respectively
$$H^{2}=\frac{\kappa^{2}}{3}\Bigg[\big(1+f(G)\big)\Big(g^{\gamma\delta}F_{0\gamma}F_{0\delta}-\frac{1}{4}
g_{00}F_{\gamma\delta}F^{\gamma\delta}\Big)+6\Big(f'(G)\big(\dot{H}H^{2}+H^{4}\big)$$
\begin{equation}
-24H^{3}\big(\ddot{H}H^{2}+2H\dot{H}^{2}+4\dot{H}
H^{3}\big)f''(G)\Big)F_{\gamma\delta}F^{\gamma\delta}-6H^{3}f'(G)\frac{\partial}{\partial
t} \big(F_{\gamma\delta}F^{\gamma\delta}\big)\Bigg],
 \end{equation}
 and
 \begin{eqnarray}
 2\dot{H}+3H^{2}&=&\kappa^{2}\Bigg[\frac{1}{12}\big(1+f(G)\big)F_{\gamma\delta}F^{\gamma\delta}+
 \Bigg(6f'(G)\big(\dot{H}H^{2}+H^{4}\big)\nonumber\\
 &-&48\Big[f''(G)\big(8\ddot{H}\dot{H}H^{3}+6\dot{H}^{3}H^{2}
 +24 \dot{H}^{2}H^{4}+6\ddot{H}H^{5}+8\dot{H}^{2}H^{6}+\dddot{H}H^{4}\big)\nonumber\\
 &+&24f'''(G)H^{2}
 \big(\ddot{H}H^{2}+2H\dot{H}^{2}+4\dot{H}H^{3}\big)^{2}\Big]
 \Bigg)F_{\gamma\delta}F^{\gamma\delta}-4
 \Big[f'(G)\big(H\dot{H}+H^{3}\big)\nonumber\\&+&24f''(G)\big(\ddot{H}H^{4}+2\dot{H}^{2}H^{3}+4\dot{H}H^{5}\big)
 \Big]\frac{\partial}{\partial t}\big(F_{\gamma\delta}F^{\gamma\delta}\big)-2f'(G)H^{2}\frac{\partial^{2}}
 {\partial
 t^{2}}\big(F_{\gamma\delta}F^{\gamma\delta}\big)\Bigg],\nonumber\\
\end{eqnarray}
 where we have neglected the second order spatial derivative of the quadratic quantity
  $F_{\gamma\delta}F^{\gamma\delta}$.\\
  Now we are going to use the following relations \cite{c19},
\begin{equation}
 g^{\gamma\delta}F_{0\gamma}F_{0\delta}-\frac{1}{4}
g_{00}F_{\gamma\delta}F^{\gamma\delta}=\frac{1}{2}\Big(|E_{i}(t,\vec{x})|^{2}+|B_{i}(t,\vec{x})|^{2}\Big),
 \end{equation}
 \begin{equation}
   F_{\gamma\delta}F^{\gamma\delta}=2\Big(|B_{i}(t,\vec{x})|^{2}-|E_{i}(t,\vec{x})|^{2}\Big).
 \end{equation}
 Here $E_{i}(t,\vec{x})$ and $B_{i}(t,\vec{x})$ are proper electric and magnetic fields respectively.
 The amplitude of the proper electric and magnetic fields on a comoving scale $L=\frac{2\pi}{k}$ with
 the comoving wave
 number $k$, are given by
\begin{eqnarray}
|E_{i}(t)|^{2}=\frac{|E_{0}|^{2}}{|(1+f(G))|^{2}a^{4}},
\,\,\,\,\,\,\, |B_{i}(t)|^{2}=\frac{|B_{0}|^{2}}{a^{4}},
\end{eqnarray}
where $|E_{0}|$ and $|B_{0}|$ are constants. Furthermore, using (12)
in (11), one obtains
\begin{eqnarray}
\frac{\partial}{\partial
t}\Big(F_{\gamma\delta}F^{\gamma\delta}\Big)&=&8\Big\{-H|B_{i}(t)|^{2}\nonumber\\
&+&\Big[H+12\frac{f'(G)}{1+f(G)}(\ddot{H}H^{2}+2\dot{H}^{2}H+4\dot{H}H^{3})\Big]|E_{i}(t)|^{2}\Big\}.
\end{eqnarray}
Because our interest is the generation of large scale magnetic
fields instead of electric fields, we neglect terms in electric
fields from this point. In this case substituting (12) and (13) in
(8) and (9) lead to
\begin{eqnarray}
H^{2}&=&\kappa^{2}\Big[\frac{1}{6}(1+f(G))+2f'(G)(\dot{H}H^{2}+9H^{4})\nonumber\\
&-&48(\ddot{H}H^{5}
+2H^{4}\dot{H}^{2}+4H^{6}\dot{H})f''(G)\Big]\frac{|B_{0}|^{2}}{a^{4}},
\end{eqnarray}
and
$$2\dot{H}+3H^{2}=\kappa^{2}\Big[\frac{1}{6}(1+f(G))+20f'(G)(3\dot{H}H^{2}-H^{4})$$
$$+48f''(G)(-8\ddot{H}\dot{H}H^{3}-6\dot{H}^{3}H^{2}+8\dot{H}^{2}H^{4}+10\ddot{H}H^{5}-8\dot{H}^{2}H^{6}+64
\dot{H}H^{6}-\dddot{H}H^{4})$$
\begin{equation}
+24f'''(G)H^{2}(\ddot{H}H^{2}+2\dot{H}^{2}H+4\dot{H}H^{3})^{2}\Big]\frac{|B_{0}|^{2}}{a^{4}},
\end{equation}
respectively, where we have used $\frac{\partial^{2}}{\partial
t^{2}}(F_{\gamma\delta}F^{\gamma\delta})\approx-8\dot{H}|B_{i}(t)|^{2}+32H^{2}|B_{i}(t)|^{2}$.\\
From equations (14) and (15), we have
$$\dot{H}+H^{2}=\kappa^{2}\Big[f'(G)(29\dot{H}H^{2}-H^{4})$$
$$+24f''(G)(-8\ddot{H}\dot{H}H^{3}-6\dot{H}^{3}H^{2}+10\dot{H}^{2}H^{4}+11\ddot{H}H^{5}-8\dot{H}^{2}H^{6}+68
\dot{H}H^{6}-\dddot{H}H^{4})$$
\begin{equation}
+12f'''(G)H^{2}(\ddot{H}H^{2}2\dot{H}^{2}H+4\dot{H}H^{3})^{2}\Big]\frac{|B_{0}|^{2}}{a^{4}}.
\end{equation}
 We examine the following
function for $f(G)$ which has been proposed in Ref. \cite{c29} as a
realistic case for both inflation and late-time acceleration,
\begin{equation}
f(G)=\frac{G^{n}}{c_{1}G^{n}+c_{2}}
\end{equation}
where  $c_{1}$ and $c_{2}$ are constants and $n$ is a positive integer.\\
For exploring power-law inflation, we assume $a=a_{0}t^{h_{0}}$,
therefore
\begin{equation}
H=\frac{h_{0}}{t},\,\,\dot{H}=-\frac{h_{0}}{t^{2}},\,\,\ddot{H}=
\frac{2h_{0}}{t^{3}},\,\,\dddot{H}=-\frac{6h_{0}}{t^{4}}.
\end{equation}
Also we use the following approximate relations which satisfy at the
inflationary period
\begin{equation}
f(G)\approx \frac{1}{c_{1}}\big(1-\frac{c_{2}}{c_{1}}G^{-n}\big),
\end{equation}
\begin{equation}
f'(G)\approx \frac{n c_{2}}{c_{1}^{2}}G^{-(n+1)},
\end{equation}
\begin{equation}
f''(G)\approx \frac{-n(n+1) c_{2}}{c_{1}^{2}}G^{-(n+2)},
\end{equation}
\begin{equation}
f'''(G)\approx \frac{n(n+1)(n+2) c_{2}}{c_{1}^{2}}G^{-(n+3)}.
\end{equation}
 By substituting above approximate relations for $f(G)$ and it's derivatives and (18) in (16),
 one can obtain
 \begin{equation}
 h_{0}=\frac{2n+1}{2}.
 \end{equation}
 If $n\gg 1$, $h_{0}$ becomes much larger than unity and power-law inflation can occur.
 Therefore, non- minimally coupled electromagnetic field with $f(G)$ gravity can be a source of inflation.
 This result is the same as in non- minimally coupled Maxwell field with $f(R)$ gravity \cite{c19}.
 We note that, in this paper we considered only the case in which the values of the terms proportional
 to $f'(G)$, $f''(G)$ and $f'''(G)$ in equations (14) and (15) are dominant to the values of the term
 proportional to $\big(1+f(G)\big)$. Because in the opposite case i.e. if the term proportional to
 $\big(1+f(G)\big)$ is dominant to the other terms, the power- low inflation can not be realized.

\section{Inflation and late-time acceleration in modified Gauss-
Bonnet gravity} Now, we consider a non- minimally coupled
electromagnetic field in a modified Gauss- Bonnet gravity proposed
in Ref. \cite{c27}.\\
We describe the model by the following action
\begin{equation}
S=\int d^{4}x
\sqrt{-g}\Big[\frac{1}{2\kappa^{2}}\big(R+F(G)\big)-\frac{1}{4}F_{\mu\nu}F^{\mu\nu}
-\frac{1}{4}f(G)F_{\mu\nu}F^{\mu\nu}\Big].
\end{equation}
We note that in this case $F(G)$ is the modified part of gravity and
it is different from the $f(G)$ in the last term in action (1). By
choosing FRW metric (5), the $(0,0)$ component and sum of $(i,i)$
components of equation of motion for $g_{\mu\nu}$, have the
following forms
\begin{equation}
H^{2}-\frac{1}{6}(GF'(G)-F(G))+4H^{3}\dot{G}F''(G)=\frac{\kappa^{2}}{3}T_{00}^{eff},
\end{equation}
and
\begin{equation}
2\dot{H}+3H^{2}+\frac{1}{2}(GF'(G)-F(G))-4H^{2}(\ddot{G}F''(G)+\dot{G}^{2}F'''(G))=-\kappa^{2}T_{ii}^{eff},
\end{equation}
where $T_{\mu\nu}^{eff}$ and $G$ are given by equations (4) and (7)
respectively. As the previous section, we neglect the contribution
of electric field and spatial derivatives of $F_{\mu\nu}F^{\mu\nu}$.
Therefore, from equations (25) and (26), one can obtain
\begin{eqnarray}
\dot{H}&+&H^{2}+\frac{1}{3}(GF'(G)-F(G))-2H^{2}(\ddot{G}F''(G)+\dot{G}^{2}F'''(G))-4H^{3}\dot{G}F''(G)\nonumber\\
&=&
\kappa^{2}\Big\{f'(G)(29\dot{H}H^{2}-H^{4})
+24f''(G)\big[-8\ddot{H}\dot{H}H^{3}-6\dot{H}^{3}H^{2}+10\dot{H}^{2}H^{4}+11\ddot{H}H^{5}\nonumber\\
&-&8\dot{H}^{2}H^{6}+68 \dot{H}H^{6}-\dddot{H}H^{4}\big]
+12f'''(G)H^{2}(\ddot{H}H^{2}+2\dot{H}^{2}H+4\dot{H}H^{3})^{2}\Big\}\frac{|B_{0}|^{2}}{a^{4}}.
\end{eqnarray}
Here, we take $F(G)$ from Ref. \cite{c27},
\begin{equation}
F(G)=\frac{(G-G_{0})^{2n+1}+G_{0}^{2n+1}}{c_{3}+c_{4}\big((G-G_{0})^{2n+1}+G_{0}^{2n+1}\big)},
\end{equation}
where $c_{3}$, $c_{4}$ are constants and $n$ is a positive integer.
$G_{0}$ corresponds to the present value of the Gauss-Bonnet
invariant. Since $F'(G)=0$ when $G=G_{0}$ and $G=\infty$, $F(G)$ can
be regarded as an effective cosmological constant. We may consider
$F(\infty)$ as the cosmological constant for the inflationary stage
and $F(G_{0})$ as that at the present time
\begin{equation}
\lim_{G\rightarrow\infty}F(G)=\frac{1}{c_{4}}=\Lambda,
\end{equation}
\begin{equation}
F(G_{0})=\frac{G_{0}^{2n+1}}{c_{3}+c_{4}G_{0}^{2n+1}}=2G_{0}.
\end{equation}
From the above equations, we find
\begin{equation}
c_{3}=\frac{G_{0}^{2n}}{2}-\frac{G_{0}^{2n+1}}{\Lambda}\approx\frac{G_{0}^{2n}}{2},\,\,\,\,\,\,\,\,\,\,\,
c_{4}=\frac{1}{\Lambda},
\end{equation}
because $\frac{G_{0}}{\Lambda}\ll1$.\\
Also, $f(G)$ is given by
\begin{equation}
f(G)=-\frac{(G-G_{0})^{2m+1}+G_{0}^{2m+1}}{c_{5}+c_{6}\big((G-G_{0})^{2m+1}+G_{0}^{2m+1}\big)},
\end{equation}
where $c_{5}$, $c_{6}$ are constants and $m$ is a positive integer.
For the inflationary epoch we can use the following approximate
relations:
\begin{equation}
F(G)\approx\frac{1}{c_{4}}\Big[1-\frac{c_{3}}{c_{4}}(G)^{-(2n+1)}\Big],
\end{equation}
and
\begin{equation}
f(G)\approx-\frac{1}{c_{6}}\Big[1-\frac{c_{5}}{c_{6}}(G)^{-(2m+1)}\Big].
\end{equation}
Because $G\rightarrow\infty$ at the inflationary stage and also
$\lim_{G\rightarrow\infty}F(G)=\Lambda$ and
$\lim_{G\rightarrow\infty}f(G)=const$, Eqs. (27) are reduced to
\begin{equation}
\dot{H}+H^{2}=\frac{\Lambda}{3}.
\end{equation}
It follows from above equation that
\begin{equation}
a(t)\propto \exp(\frac{\Lambda}{3})^{\frac{1}{2}}t,
\end{equation}
so that exponential inflation can be realized. Thus, we conclude
that the terms in $F(G)$ on the left hand side of Eqs. (27) can be a
source of inflation, in addition to $f(G)$ on the right hand side of
Eqs. (27).\\
At the present time, because $G-G_{0}\ll1$, if $m>n$, $f(G)$ becomes
constant more rapidly than $F(G)$ in the limit $G\rightarrow G_{0}$.
For such a case, when $G\rightarrow G_{0}$ Eqs. (27) lead to
\begin{equation}
\dot{H}+H^{2}=\frac{2G_{0}}{3}.
\end{equation}
Then, from this equation one can obtain
\begin{equation}
a(t)\propto \exp(\frac{2G_{0}}{3})^{\frac{1}{2}}t,
\end{equation}
so that the late time acceleration of the universe can be realized.
Thus, these results are in agreement with the results of Ref.
\cite{c19} where non-minimal Maxwell-F(R) gravity has been
considered.

\section{Conclusion}
To summarize, the non-minimal gravitational coupling of
electromagnetic field with Gauss-Bonnet invariant function, $f(G)$,
has been considered in Friedmann-Robertson-Walker background metric.
Such a non-minimal coupling has been examined in the framework of
general relativity. We have shown that power-law inflation can be
realized in this model which is described by action (1). We have
also studied cosmology in non-minimally coupled electromagnetic
field in the framework of modified Gauss-Bonnet gravity, $F(G)$. It
has been shown that both inflation and late-time acceleration of the
universe can be realized in such a model proposed in Ref.
\cite{c27}.

\end{document}